## OPTICS
## AND LASER PHYSICS

# Synchronous Comparison of Two Thulium Optical Clocks

**A. Golovizin[a,b,*], D. Mishin[a], D. Provorchenko[a], D. Tregubov[a], and N. Kolachevsky[a,b]**

[a] *Lebedev Physical Institute, Russian Academy of Sciences, Moscow, 119991 Russia*
[b] *International Center of Quantum Technologies, Moscow, 121205 Russia*
*\*e-mail: artem.golovizin@gmail.com*

The experimental comparison of two thulium optical lattice clocks in a time interval of up to one hour has been carried out. The synchronous comparison of a clock transition in two independent atomic ensembles using a single ultrastable laser has allowed us to eliminate fluctuations of the laser frequency from the measured frequency difference and to reach a relative measurement error of $10^{-16}$ after 500-s averaging, which corresponds to a relative instability of $2 \times 10^{-15}/\sqrt{\tau}$. The successful demonstration of the long-term operation of two systems using the synchronous comparison of clock transitions opens the possibility of studying systematic shifts in thulium optical clocks with an uncertainty of $10^{-17}$.

## INTRODUCTION

Optical clocks currently demonstrate the lowest systematic frequency uncertainty at a level of $10^{-18}$ [1–3]. The possibility of measuring the frequency difference of atomic transition with a relative inaccuracy of $10^{-18}$ to $10^{-21}$, which in particular opens up the possibility of studying the gravitational redshift on a millimeter scale [4–6]. Advances in the development of optical clocks led to the understanding of the necessity to redefine second as the SI unit of time, which is scheduled for 2030 [7]. To implement the unit of time based on optical clocks, it is necessary to accurately compare times and frequencies between different metrological institutions with a measurement accuracy of about $10^{-18}$. This can be done using both a network of stabilized optical fibers [8, 9] and transportable optical clocks [10–13] with the appropriate characteristics. Transportable optical clocks are used for intercontinental comparison of optical clocks, for relativistic geodesy [14], and for the space-based synchronization (navigation, data transmission).

Thulium optical clocks are promising for building transportable systems due to the low sensitivity of the clock transition frequency at a wavelength of 1140 nm to the environment. In particular, our group previously demonstrated a record low clock transition frequency shift induced by thermal radiation [15] and a low sensitivity to the magnetic field at the formation of the synthetic frequency [16]. Moreover, thulium optical clock has convenient optical wavelengths for laser cooling, trapping and clock transition spectroscopy [15, 17].

In this work, we report the first experimental results on the synchronous comparison of two thulium optical clocks. In these experiments, the synchronous comparison technique was implemented; i.e., clock transitions in two systems (optical clocks) were simultaneously excited using a single ultrastable clock laser. This allowed us to exclude the influence of phase noise of the clock laser, in particular, due to the periodic interrogation of the clock transition (Dicke effect), from the measured frequency difference and to achieve instability limited by the standard quantum limit for independent particles [4, 6].

## EXPERIMENTAL SETUP
## AND THE MEASUREMENT PROTOCOL

The scheme of thulium atomic levels involved in the experiment is presented in Fig. 1a. Operation principles of the primary and secondary magneto-optical traps, optical pump, and simultaneously excitation of two 4–3 and 3–2 clock transitions (see Fig. 1a), as well as the measurement of the efficiency of the excitation of the clock transition, were described in detail in [16–21]. All laser systems, including the ultrastable clock laser, are common, whereas the designs of two vacuum systems (main [16] and compact [22]), where ultracold thulium atoms are trapped, are somewhat different. We note that a simplified slowing scheme of atoms and an intravacuum cavity for the optical lattice were implemented in the compact setup (fabricated in 2021) smaller than $80 \times 50 \times 60$-cm. It was mounted on a single optical plate and all necessary optical radiation was guided using single-mode optical fibers.

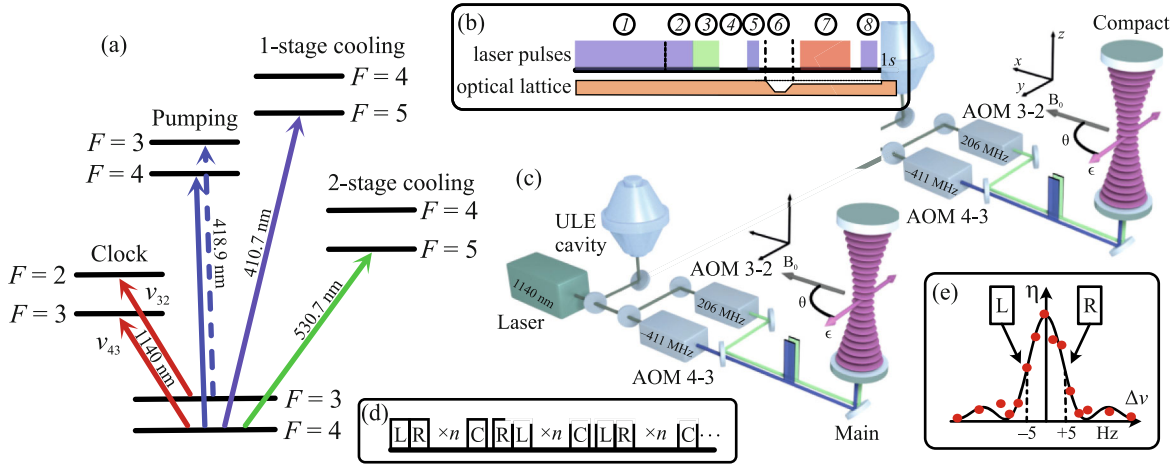

**Fig. 1.** (Color online) Sketch of the experiment. (a) Scheme of the energy levels of the thulium atom. The frequency of the probe clock radiation is stabilized to the frequencies $v_{43}$ and $v_{32}$ of the 4–3 and 3–2 transitions. (b) Single measurement cycle: (*1*) Zeeman slowing, (*2*) cooling in the primary magneto-optical trap, (*3*) cooling in the secondary magneto-optical trap, (*4*) application of the magnetic field and expansion of atoms untrapped in the optical lattice, (*5*) optical pumping, (*6*) filtration of hot atoms, (*7*) excitation of clock transitions, (*8*) readout of the populations of clock levels. (c) Layout of the experimental setup for the formation of clock pulses and interrogation of atoms confined in the optical lattice in the main and compact systems: (AOM) acousto-optic modulator and (ULE cavity) ultrastable reference cavity. (d) Sequence of measurement cycles on the left (L) and right (R) slopes and calibration measurements (C). A calibration measurement is carried out after *n* pairs of measurements on the left and right slopes. (e) (Red circles) Measured contour of the excitation line of the 1440-nm clock transition and (solid line) its theoretical approximation by the sinc function with the indicated interrogation points on the left and right slopes.

This will allow one to raise the system at a height of about 1 m to study the gravitational frequency shift.

As shown in our previous works [16, 23], the effect of the magnetic field is excluded due to the equality of the Zeeman shifts of the lower and upper hyperfine sublevels in absolute value at the formation of the synthetic frequency

$$v_s = \frac{v_{32} + v_{43}}{2}. \qquad (1)$$

The differential frequency

$$v_d = v_{32} - v_{43} \qquad (2)$$

lies in the radiofrequency band and reflects both the influence of fluctuations of the external magnetic field and noise introduced by the atomic ensemble (e.g., projective noise) and by the readout system (noise of photodetectors and electronics).

A single measurement cycle took $T_c = 1$ s and consisted of stages schematically shown in Fig. 1b and were described in detail in [21, 24] and in the supplementary material. Such a cycle was repeated during the entire measurement run (see Fig. 1d). Two systems were synchronized at the end of Zeeman slowing (indicated by the dotted line in Fig. 1b) in each cycle, which ensured the simultaneous beginning and end probe clock pulses with an accuracy of about 1 μs. To measure the frequency detuning of the clock laser from atomic transitions and to stabilize the radiation frequency to each transition, the clock transition in both systems was excited at the detuning of the probe radiation $\pm \Delta v_c/2 = \pm 5$ Hz from the current resonant frequency (alternately on the right and left slopes of the line contour, see Fig. 1e), where $\Delta v_c = 10$ Hz is the width of the line contour at the 80-ms probe clock pulse. The frequency detuning of laser radiation from the resonance is determined by $\delta^{err} = -\Delta v_c(\eta^+ - \eta^-)$, where $\eta^+(\eta^-)$ are the measured excitation efficiency on the right (left) slope. The introduced correction to the frequency $\Delta v^{corr}$ of the corresponding acousto-optic modulator, which compensates the frequency detuning of the clock laser from the atomic transition, was calculated using a digital proportional-integral controller with the typical coefficients $P = 0.15$ and $I = 0.1$ [23]. Calibrations measurements were conducted periodically (once per *n* cycles of "main" measurements), in particular, to determine the angle between the polarization vector of the lattice and the driven magnetic field [16].

Thus, the array

$$v^{corr}[n] = \sum_{i=0}^{n} \Delta v^{corr}[i], \qquad (3)$$

which is the accumulated correction to the frequency of the corresponding acousto-optic modulator after *n* cycles of measurements, was formed for each clock transition in each system (main and compact), and the array

$$\Delta v^{meas}[n] = v^0 + v^{corr}[n] + \delta^{err}[n] \qquad (4)$$

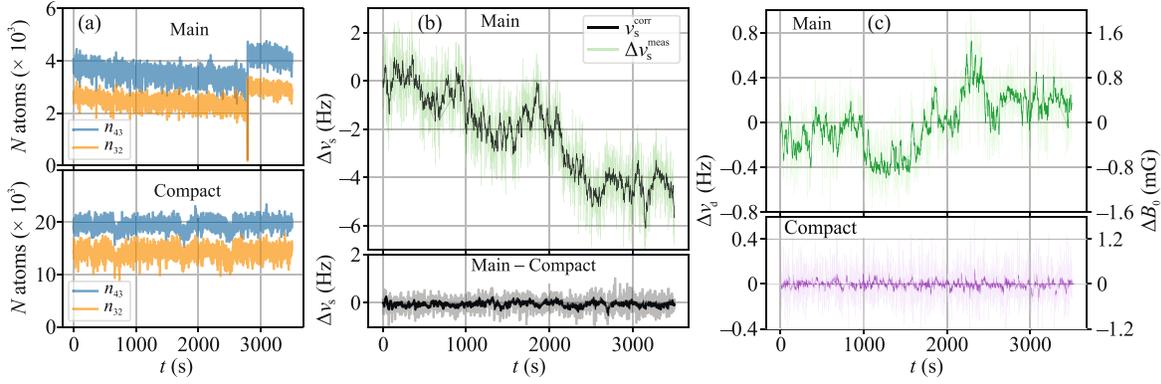

**Fig. 2.** (Color online) (a) Number of atoms in the (upper part) main and (lower part) compact systems. The initial numbers of atoms on $|g, F = 4, m_F = 0\rangle$ (4–3 clock transition) and on $|g, F = 4, m_F = 0\rangle$ (3–2 clock transition) are given in blue and orange, respectively. (b) (Upper part) Light green line is the measured frequency deviation of laser radiation from the clock frequency in the main system $\Delta\nu_s^{meas}$ and the black line is the averaged value $\nu_s^{corr}$ corresponding to the introduced frequency correction frequency of (see Eq. (5)). (Lower part) Gray and black lines are the measured and averaged differences of the synthetic clock frequencies between the compact and main systems, respectively. (c) Time dependences of the (left axis) differential frequency and (right axis) driven magnetic field in the (upper part) main and (lower part) compact systems. The light and dark lines correspond to $\Delta\nu_d^{meas}$ and $\nu_d^{corr}$, respectively.

reflects the instantaneous frequency detuning of the clock radiation from the atomic transition because the error $\delta^{err}[n]$ was measured at the frequency of the acousto-optic modulator corrected by $\nu^{corr}[n]$, and $\nu^0$ is the initial frequency shift of radiation by the corresponding acousto-optic modulator.

Arrays of corrections and measured detunings of the synthetic and differential frequencies for each optical system can be expressed in terms of these quantities

$$\nu_s^{corr} = \frac{\nu_{32}^{corr} + \nu_{43}^{corr}}{2}, \quad \nu_d^{corr} = \nu_{32}^{corr} - \nu_{43}^{corr};$$

$$\Delta\nu_s^{meas} = \frac{\Delta\nu_{32}^{meas} + \Delta\nu_{43}^{meas}}{2}, \quad (5)$$

$$\Delta\nu_d^{meas} = \Delta\nu_{32}^{meas} - \Delta\nu_{43}^{meas}.$$

Thus, $\Delta\nu_d^{meas}$ reflects the measured differential frequency $\nu_d$ in a single system of thulium optical clocks (all radio frequencies are determined compared to an active hydrogen maser). The frequency difference between two systems is $\Delta\nu_x^{meas,m} - \Delta\nu_x^{meas,c}$, where the subscript $x$ specifies the 4–3, 3–2 transition frequencies or the synthetic frequency ($x = s$). Below, this quantity is used to characterize the instability of the difference between clock frequencies of two optical clocks.

## RESULTS

Figures 2 and 3 present the results of the simultaneous comparison of two systems. The lattice depth in the main/compact system was $300E_r/500E_r$ (where $E_r = 1$ kHz is the recoil energy of the lattice photon) at a driven magnetic field of 500/340 mG. The entire experiment took 1 h. Figure 2a shows the long-term dynamics of the number of atoms during the comparison on each of the clock transitions in the main and compact systems in the process of experiment. The number of atoms in the main system decreased in the first 2800 s; after that, it vanished abruptly and recovered to the initial value in a short time. Such a behavior was due to the temperature drift of the amplifying cavity of the optical lattice and to the feedback system, which led to overmatching. We are going to upgrade the amplifying cavity to eliminate this effect.

Because of a feature of optical pumping, the number of atoms in the initial state $|g, F = 4, m_F = 0\rangle$ is approximately 25% higher than that in the state $|g, F = 3, m_F = 0\rangle$. Short-term fluctuations of the number of atoms in both systems were at a level of 5–8% (one standard deviation. Since the numbers of excited and unexcited atoms were measured in each measurement cycle and the excitation efficiencies were calculated independently for each clock transition, the difference of the initial populations and fluctuations of the number of atoms insignificantly affected the tuning of the laser radiation frequency to each of the 4–3 and 3–2 clock transitions.

The light green and black lines in the upper part of Fig. 2b are the time dependences of the measured deviation $\Delta\nu_s^{meas}$ from the initial frequency shift and the introduced correction $\nu_s^{corr}$, respectively, for the synthetic clock frequency in the main system. Instantaneous fluctuations of the frequency of the clock laser, which can reach 2 Hz, are less than the half-width of the transition line $\Delta\nu_c/2 = 5$ Hz, which ensures the correct determination of the instantaneous frequency error. The gray and black lines in the lower part of Fig. 2b are the difference of the measured values minus the initial frequency shift and the introduced corrections of the synthetic frequencies in two systems, respectively. Fluctuations of the instantaneous difference of the synthetic clock frequencies in two systems are much smaller than fluctuations of the shift of individual synthetic frequencies because the interrogation of the clock transitions in two systems is carried out synchronously, which makes fluctuations of the clock laser frequency common for two systems.

The stability of the driven magnetic field $B_0$ can be characterized by the differential frequency $\nu_d = \nu_{32} - \nu_{43}$, which has the quadratic Zeeman coefficient $\beta_d \approx 514$ Hz/G$^2$. The light and dark lines in Fig. 2c are the time dependences of the (left axis) differential frequencies $\Delta\nu_d^{meas}$ and $\nu_d^{corr}$ (see Eq. (5), respectively, in the (upper part) main and (lower part) compact systems. Using the formula for the quadratic Zeeman coefficient, one can recalculate frequency fluctuations to fluctuations of the magnetic field $\Delta B_0 = \Delta\nu_d/(2\beta_d B_0)$ (see the right axis in Fig. 2c). The difference of results for two systems is due to the upgraded design of the compact system. Fluctuations of the laboratory magnetic field in the direction perpendicular to the quantization axis affect the shift induced by the optical lattice due to the nonzero differential tensor polarizability of the clock transitions [16]. However, even for the amplitude of fluctuations observed in the main system, the corresponding frequency shift is less than 1 mHz, which is much smaller than the current measurement error.

Allan deviations for the difference in the synthetic clock frequency $\nu_s^{meas}$ between the main and compact systems, as well as the difference in the frequencies $\nu_{43}$ and $\nu_{32}$ of the 4–3 and 3–2 transitions, respectively, between two systems are shown in Fig. 3. It is seen that the instability of the frequencies $\nu_{43}$ and $\nu_{32}$ begins to increase after $\tau \approx 100$ s, which is primarily due to fluctuations of the magnetic field in the main system. At the same time, the Allan deviation for the synthetic frequency $\nu_s^{meas}$, which is insensitive to the magnetic field, continues to decrease and reaches $10^{-16}$ after 500-s measurements. The black solid line is the deviation given by the formula $\sigma_y = 2 \times 10^{-15}/\sqrt{\tau}$, where $\tau$ is the measurement time in seconds, which is an approximation for $\nu_s^{meas}$ at long averaging times and corresponds to the results obtained in other world laboratories.

Figure 3 also presents the Allan deviations of the differential frequencies $\Delta\nu_d^{meas}$ in two systems normal-ized to the optical transition frequency. As mentioned above, fluctuations $\Delta\nu_d^{meas}$ are due to noise in the magnetic field, as well as to projection and readout noise. The deviation $\Delta\nu_d^{meas}$ in the compact system after 200-s averaging is less than $2 \times 10^{-17}$ (violet points). This result is very hopeful because it confirms that a significant part of noise sources in our compact optical clock makes a contribution less than $2 \times 10^{-17}$, which makes it possible to expect the corresponding comparison accuracy in future. At the same time, this contribution in the main system (green points) is $(3-8) \times 10^{-16}$, which is more than an order of magnitude larger. As mentioned above, this is due to higher fluctuations of the magnetic field and to a smaller number of atoms.

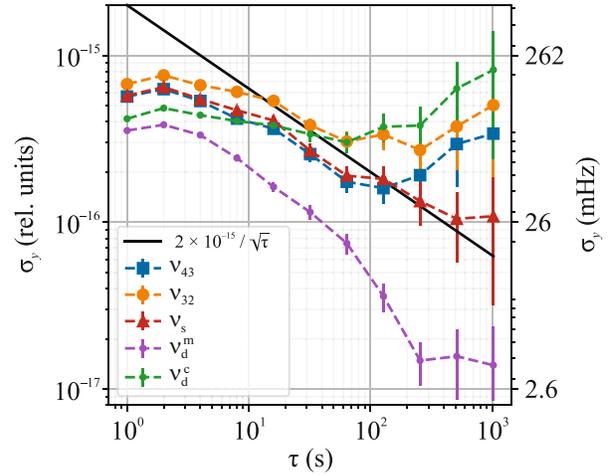

**Fig. 3.** (Color online) Allan deviation of the frequency difference between two thulium optical clocks according to synchronous comparison: (orange circles/blue squares) the deviation of the measured frequency differences $\Delta\nu_{43}^{meas}/\Delta\nu_{32}^{meas}$ of the 4–3/3–2 clock transition in the main and compact systems and (red triangles) the deviation for the synthetic frequency $\Delta\nu_s^{meas}$. The increase in the Allan deviation for the frequencies $\nu_{43}$ and $\nu_{32}$ at times longer than 100 s is primarily due to fluctuations of the magnetic field in the main system. The green (violet) points demonstrate the Allan deviation of the differential frequency $\Delta\nu_d^{meas}$ with respect to an active hydrogen maser ($\nu_d \approx 617$ MHz) normalized to the optical frequency of the clock transition in the main (compact) system.

In turn, the observed instability of the difference in the synthetic frequency $\nu_s^{meas}$ between two systems (red triangles) is also at a level of $10^{-16}$, which can be explained by the contribution from uncorrelated noise between two systems. In particular, it can be projection noise of the interrogated atomic ensemble in the main system and frequency fluctuations introduced by optical fibers through which radiation of the clock laser is delivered to each system. The last fluctuations can be eliminated by the active compensation [25], which is planned in the near future.

## CONCLUSIONS

To summarize, we have reported the results of the first long-term comparison of two thulium optical clocks with 1-h measurements. The relative instability of the frequencies of two systems has reached $10^{-16}$ after 500 s of measurement, which is comparable with a number of foreign samples. We note that this value is extreme for microwave standards (active hydrogen masers, rubidium and cesium fountains) but its achievement requires averaging intervals longer than a day. To eliminate the effect of noise in the clock laser and the Dicke effect, we have used the synchronous interrogation method, where clock transitions are excited simultaneously in both systems. This has allowed us to reduce the instability of the measured frequency per second from $\sigma_L = 0.56$ Hz observed in an individual system to the value $\sigma_s = 0.15$ Hz for the difference between two synthetic frequencies. We note that the latter value is below the instability of the clock laser frequency $\sigma_L \approx 0.26$ Hz per second [26]. The reached instability of the frequency is more than a factor of 6 lower than that in measurements in a single system with the successive change of the parameters (e.g., the magnetic field) to determine their effect on the frequency [16]. We have also demonstrated that fluctuations of the magnetic field, which were about 1 mG in the main system, do not affect the synthetic clock frequency.

In turn, studying the differential frequency between two components of the clock transition, we estimated the contribution of intrinsic noise in the compact system as less than $2 \times 10^{-17}$, which indicates that such an accuracy level can be reached when comparing two systems after some technical improvements.

It is noteworthy that the reached accuracy $10^{-16}$ of the measurement of the difference between the synthetic clock frequencies is already enough to observe the gravitational shift at a change in the relative height of atomic ensemble by 1 m. An increase in the reliability of individual laser systems and the amplifying cavity in the main setup together with the compensation of noise in optical fibers will allow longer measurements with a lower instability of the frequency comparison in the near future.

## SUPPLEMENTARY INFORMATION

The online version contains supplementary material available at http://doi.org/10.31857/S0021364024600873.

## FUNDING

This work was supported by the Russian Science Foundation (project no. 21-72-10108).

## CONFLICT OF INTEREST

## OPEN ACCESS

*Translated by R. Tyapaev*